\documentclass[allclo]{FBSart}
\usepackage{epsfig}
\newcommand{\cd}{\makebox[0.08cm]{$\cdot$}}
\title{Higher Fock sectors in Wick-Cutkosky model\footnote{Presented
at Light-Cone 2004, Amsterdam, 16 - 20 August}}
\author{V.A. Karmanov$^a$ and Dae Sung Hwang$^b$}
\institute{$^a$ Lebedev Physical Institute, Leninsky Prospekt 53,
119991 Moscow, Russia\\
$^b$ Department of Physics, Sejong University,
Seoul 143--747,
Korea}
\runningauthor{V.A. Karmanov and Dae Sung Hwang}
\runningtitle{LC 2004}
\begin{document}
\maketitle
\begin{abstract}
In the Wick-Cutkosky model we analyze nonperturbatively, in light-front
dynamics, the contributions of two-body and higher Fock sectors to the
total norm and electromagnetic form factor. It turns out that two- and
three-body sectors always dominate. For maximal value of coupling constant
$\alpha=2\pi$, corresponding to zero bound state mass $M=0$, they
contribute 90\% to the norm. With decrease of $\alpha$ the two-body
contribution increases up to 100\%. The  form factor asymptotic is always
determined by two-body  sector.
\end{abstract}
\section{Introduction}
In field theory, the state vector  $\left\vert p \right>$ is described by
an infinite set of the Fock components, corresponding to different numbers
of particles. In light-front dynamics  \cite{cdkm,bpp} the state vector is
defined on the light front  plane $\omega\cd x=0$, where  $\omega$ is the
null four-vector ($\omega^2=0$). The wave functions are expressed in terms
of the variables $\vec{k}_{\perp},x$:
$\psi=\psi(\vec{k}_{1\perp},x_1;\vec{k}_{2\perp},x_2;
\ldots;\vec{k}_{n\perp},x_n)$. The total norm (equaled to 1) is given by
sum over all the sectors:  $\sum_n N_n=1$, where $n$-body contribution
$N_n$ reads:
\begin{eqnarray}\label{norm}
N_n&=&(2\pi)^3\int |\psi(\vec{k}_{1\perp},x_1;\vec{k}_{2\perp},x_2;
\ldots;\vec{k}_{n\perp},x_n)|^2
\nonumber\\
&\times&
\delta^{(2)}(\sum_{i=1}^n\vec{k}_{\perp i})
\delta(\sum_{i=1}^n x_i-1)2\prod_{i=1}^n{d^2k_{\perp
i}dx_i\over(2\pi)^3 2x_i} \ .
\end{eqnarray}

In applications, the infinite set of the Fock components is usually
truncated to a few components only. The belief that a given Fock sector
dominates (with two or three quarks, for instance) is often based on
intuitive expectations and on ``experimental evidences'' rather than on
field-theoretical analysis.

In the Wick-Cutkosky model   two massive scalar particles interact by the
ladder exchange of massless scalar particles. Two-body sector contains two
massive particles. Higher sectors contain two massive and $1,2,\ldots$
massless constituents.

In the present paper, based on the work \cite{hk04}, we present the
results of our study in the Wick-Cutkosky model of contributions of the
two- and three-body sectors to the total norm. Subtracting them from 1, we
get total contribution of all the sectors with $n\ge 4$. Besides, we also
calculate their contributions to the electromagnetic form factor.
Calculations are carried out nonperturbatively in full range of binding
energy $0\le B\le 2m$.
%
\section{Bethe-Salpeter amplitude in Wick-Cutkosky model}
We use Bethe-Salpeter (BS) amplitude known explicitly in the Wick-Cutkosky
model \cite{wcm}. For the ground state with zero angular momentum it
reads:
\begin{equation}
\Phi (k,p)=-\frac{i}{\sqrt{4\pi }}\int_{-1}^{+1}
\frac{g_M(z) d z}{(m^2-M^2/4-k^2-zp\cd k-i\epsilon)^3}\ ,
\label{bs8p}
\end{equation}
where $k$ and $p$ are relative and total four-momenta, $m$ is the massive
constituent mass, $M$ is the mass of the composite system. The
representation (\ref{bs8p}) is valid and exact for the zero-mass exchange.
The function $g_M(z)$ is determined by the integral equation:
\begin{equation}\label{g0}
g_M(z)=\frac{\alpha}{2\pi}\int_{-1}^1 K(z,z')g_M(z') d z'
\end{equation}
with the kernel: $$ K(z,z')=
\frac{m^2}{m^2-\frac{1}{4}(1-{z'}^2)M^2}\left[
\frac{(1-z)}{(1-z')}\theta(z-z') +\frac{(1+z)}{(1+z')}\theta(z'-z)\right].
$$ Here $\alpha=g^2/(16\pi m^2)$ and $g$ is the coupling constant in the
interaction Hamiltonian $H^{int}=-g\varphi^2(x)\chi(x)$. In
nonrelativistic limit the interaction is reduced to the Coulomb potential
$V(r)=-\frac{\alpha}{r}$.

The normalization condition for $g_M(z)$ is found from the requirement
that the full electromagnetic form factor $F_{full}(Q^2)$ (calculated with
full state vector $\left\vert p \right>$ and, hence, incorporating all the
Fock components) equals to 1 at $Q=0$. Form factor is expressed in terms
of the BS amplitude:
\begin{eqnarray}\label{ffbs}
&&(p+p')^\mu F_{full}(Q^2) \\ &=&-i\int {d^4k\over (2\pi)^4}\
(p+p'-2k)^\mu \; (m^2-k^2)\; \Phi \left(\frac{1}{2}p
-k,p\right)\Phi  \left(\frac{1}{2}p'-k,p'\right). \nonumber
\end{eqnarray}
We substitute here the BS amplitude (\ref{bs8p}) and find the
normalization of $g_M(z)$ from the equality $F_{full}(0)=1$. The details
of calculations are given in \cite{hk04}.

The function $g_M(z)$ is found from (\ref{g0}) analytically in the
limiting cases of small binding energy $B=2m-M$ ($\alpha\to 0$, $B\to 0$,
$M\to 2 m$) and of extremely large binding energy ($\alpha=2\pi$, $B=2m$,
$M=0$). In the case $M\to 2 m$ it reads:
\begin{equation}\label{g1}
g_{M}(z)=8\sqrt{2}\pi\alpha^{5/2}
m^3\left(1+\frac{5\alpha}{\pi}\log\alpha\right)
\left[1-|z|+\frac{\alpha}{2\pi}(1 +
|z|)\log(z^2+\alpha^2/4)\right] .
\end{equation}
In contrast to the solution found in \cite{wcm}, eq. (\ref{g1}) is
calculated to the next $\alpha$ order, keeping, however, the leading
$\log\alpha$ term (i.e., neglecting $const$ relative to $\log\alpha$).

In the opposite case $M=0$ $g_{M=0}(z)$ has the form:
\begin{equation}\label{M0}
g_{M=0}(z)=6\sqrt{30}\pi^{3/2}m^3(1-z^2)\ .
\end{equation}

For arbitrary $M$ the function $g_{M}(z)$ is found from (\ref{g0})
numerically.
%
\section{Two- and three-body contributions}
Knowing the BS amplitude, we  extract from it the two-body wave function
\cite{cdkm}:
\begin{equation} \label{bs8}
\psi(\vec{k}_{\perp},x) =\frac{(\omega\cd k_1 )(\omega\cd k_2
)}{\pi (\omega\cd p)}\int_{-\infty }^{+\infty }\Phi
(k+\beta\omega,p)d\beta.
\end{equation}
This relation is independent of any model. In Wick-Cutkosky model,
substituting (\ref{bs8p}) into (\ref{bs8}), we find:
\begin{equation}\label{bs10}
\psi(\vec{k}_{\perp},x) =\frac{x(1-x)g_M(1-2x)}{2\sqrt{\pi }
\Bigl(\vec{k}_{\perp}^2+m^2-x(1-x)M^2\Bigr)^2}\ .
\end{equation}

Substituting (\ref{bs10}) into (\ref{norm}), we obtain the two-body
contribution to the full normalization:
\begin{equation}\label{norm2}
N_2={1\over 192 \, \pi^3}\int_0^1 {x(1-x)\; g_M^2(2x-1)
d x\over \Bigl(m^2-x(1-x)M^2{\Bigr)}^3} \ .
\end{equation}

\begin{figure}[ht]
\begin{center}
\psfig{figure=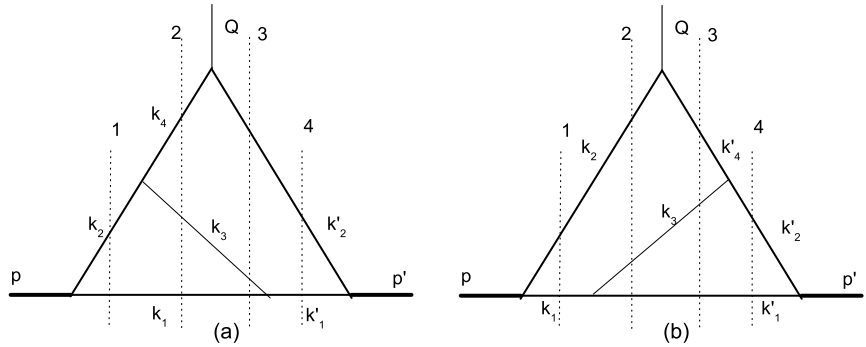}
\caption{Three-body contributions to form factor.} \label{3b}
\end{center}
\end{figure}

Three-body contribution $N_3$ is found in  \cite{hk04} by calculating the
amplitudes  Fig. \ref{3b} (at $Q=0$), where two-body vertices are
determined by the wave function (\ref{bs10}).
%
\section{Results}
For small $\alpha$, with eq. (\ref{g1}) for $g_M(z)$,  the contributions
$N_2$ and $N_3$ to the total norm are found analytically (up to $\alpha
\log \alpha$ order):
\begin{equation}\label{N2anal}
N_{2}=1-\frac{2\alpha}{\pi}\log\frac{1}{\alpha},\quad
N_{3}=\frac{2\alpha}{\pi}\log\frac{1}{\alpha},\quad N_{n\ge 4}={\cal
O}(\alpha^2).
\end{equation}

For $\alpha=2\pi$ ($B=2m$, $M=0$), with $g_M(z)$ given by  (\ref{M0}), we
get:
\begin{equation}\label{N2}
N_2=\frac{9}{14}\approx 64\%,\quad N_3\approx 26\%,\quad
N_{n\ge 4}\approx 10\%.
\end{equation}

\begin{figure}[ht]
\begin{center}
\psfig{figure=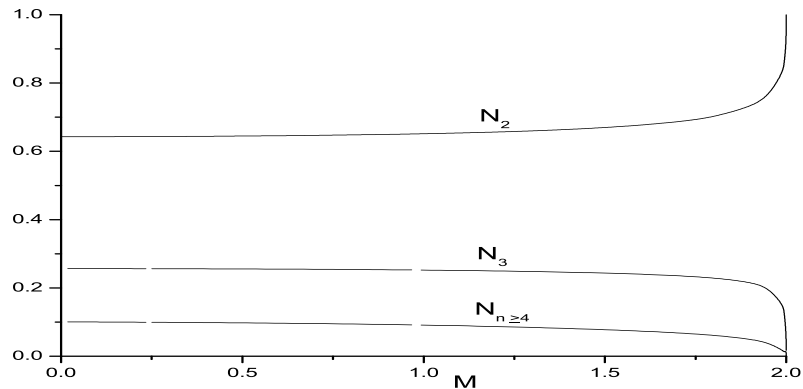,height=60mm,width=120mm} \caption{Contributions to
the total norm $N_{n=2} + N_{n=3} + N_{n\ge 4}=1$ of the Fock sectors with
the constituent numbers $n=2$, $n=3$ and  $n\ge 4$ vs. the bound state
mass $M$ (in units of $m$).} \label{N3_M}
\end{center}
\end{figure}

For  $\alpha$ in the interval $2\pi \ge \alpha \ge 0$, corresponding to $0
\le M \le 2m$, the values of $N_2$, $N_3$ and $N_{n\ge 4}$ vs. $M$ are
found numerically and they are shown in Fig. \ref{N3_M}.

We find that two-body sector always dominates. The sum $N_2+N_3$
contributes 90\% even in the extremely strong coupling case, as we see in
(\ref{N2}). This result is non-trivial, since for $\alpha=2\pi$ one might
expect just the opposite relation of the $N_2+N_3$ and $N_{n\ge 4}$
contributions.  For any $\alpha$, asymptotic behavior of the form factor
$F_{full}(Q^2)$ is determined by the two-body Fock sector \cite{hk04}.
\end{document}